\documentstyle[amssymb]{article}
\newtheorem{th}{Theorem}[section]
\newtheorem{prop}[th]{Proposition}
\newtheorem{cor}[th]{Corollary}

\newtheorem{lem}[th]{Lemma}

\begin{document}

\title{Certain basic inequalities for submanifolds in a $\left( \kappa ,\mu \right)
$-contact space form}
\author{Mukut Mani Tripathi\thanks{%
First author is Post-Doctoral Researcher under Brain Korea-21 Project at
Chonnam National University, Korea. }, Jeong-Sik Kim\thanks{%
Second and third authors would like to acknowledge financial support in part
from Korea Science and Engineering Foundation Grant (R01-2001-00003).} and
Jaedong Choi}
\date{}
\maketitle

{\bf Abstract.} Certain basic inequalities between intrinsic and extrinsic
invariants for a submanifold in a $\left( \kappa ,\mu \right) $-contact
space form are obtained. As applications we get some results for invariant
submanifolds in a $\left( \kappa ,\mu \right) $-contact space form. \medskip



\section{Introduction}

In \cite{chen93}, B.-Y. Chen established a sharp inequality for a
submanifold in a real space form involving intrinsic invariants, namely the
sectional curvatures and the scalar curvature of the submanifold; and the
main extrinsic invariant, namely the squared mean curvature. Similar
inequalities were established in case of submanifolds of Sasakian space
forms also (\cite{carriazo99},\cite{KK99}).

Recently, T. Koufogiorgos introduced the notion of $\left( \kappa ,\mu
\right) $-contact space form (\cite{koufog}), which contains the well known
class of Sasakian space forms for $\kappa =1$. Thus it is worthwhile to
study relationships between intrinsic and extrinsic invariants of
submanifolds in a $\left( \kappa ,\mu \right) $-contact space form. The
paper is organized as follows. Section~\ref{sect-prel} contains necessary
details about $\left( \kappa ,\mu \right) $-contact space form and its
submanifolds. In section~\ref{sect-ineq}, we state a Lemma relating scalar
curvature, squared mean curvature and squared second fundamental form for a
submanifold tangential to the structure vector field in a $\left( \kappa
,\mu \right) $-contact space form, then we obtain two basic inequalities
involving the scalar curvature and the sectional curvatures of the
submanifold on left hand side and the squared mean curvature on the right
hand side. In the last section, we study invariant submanifolds of $\left(
\kappa ,\mu \right) $-contact space forms. We also obtain a B.-Y. Chen
inequality for Chen like $\delta $-invariant for invariant submanifolds in a
$\left( \kappa ,\mu \right) $-contact space form.

\section{Preliminaries\label{sect-prel}}

A $\left( 2m\!+\!1\right) $-dimensional differentiable manifold $\tilde{M}$
is called an almost contact manifold if its structural group can be reduced
to $U\!\left( m\right) \!\times \!1$. Equivalently, there is an almost
contact structure $\left( \varphi ,\xi ,\eta \right) $ consisting of a $%
\left( 1,1\right) $ tensor field $\varphi $, a vector field $\xi $, and a $1$%
-form $\eta $ satisfying $\varphi ^{2}=-I+\eta \otimes \xi $ and (one of) $%
\eta (\xi )=1$, $\varphi \xi =0$, $\eta \circ \varphi =0$. The almost
contact structure is said to be {\em normal} if the induced almost complex
structure $J$ on the product manifold $\tilde{M}\times {\Bbb R}$ defined by $%
J\left( X,\lambda d/dt\right) =\left( \varphi X-\lambda \xi ,\eta \left(
X\right) d/dt\right) $ is integrable, where $X$ is tangent to $\tilde{M}$, $%
t $ the coordinate of ${\Bbb R}$ and $\lambda $ a smooth function on $\tilde{%
M}\times {\Bbb R}$. The condition for being normal is equivalent to
vanishing of the torsion tensor $\left[ \varphi ,\varphi \right] +2d\eta
\otimes \xi $, where $\left[ \varphi ,\varphi \right] $ is the Nijenhuis
tensor of $\varphi $. Let $\left\langle \,,\right\rangle $ be a compatible
Riemannian metric with $\left( \varphi ,\xi ,\eta \right) $, that is, $%
\left\langle X,Y\right\rangle $ $=$ $\left\langle \varphi X,\varphi
Y\right\rangle $ $+$ $\eta \left( X\right) \eta \left( Y\right) $ or
equivalently, $\Phi \left( X,Y\right) \equiv \left\langle X,\varphi
Y\right\rangle =-\left\langle \varphi X,Y\right\rangle $ along with $%
\left\langle X,\xi \right\rangle =\eta \left( X\right) $ for all $X,Y\in T%
\tilde{M}$. Then, $\tilde{M}$ becomes an almost contact metric manifold
equipped with an almost contact metric structure $\left( \varphi ,\xi ,\eta
,\left\langle \,,\right\rangle \right) $. An almost contact metric structure
becomes a {\em contact metric structure} if $\Phi =d\eta $. A normal contact
metric manifold is a {\em Sasakian manifold}. An almost contact metric
manifold is Sasakian{\em \ }if and only if $(\tilde{\nabla}_{X}\varphi
)Y=\left\langle X,Y\right\rangle \xi -\eta (Y)X$ for all $X,Y\in T\tilde{M}$%
, where $\tilde{\nabla}$ is Levi-Civita connection. Also, a contact metric
manifold $\tilde{M}$ is Sasakian if and only if the curvature tensor $\tilde{%
R}$ satisfies $\tilde{R}(X,Y)\xi =\eta (Y)X-\eta (X)Y$ for all $X,Y\in T%
\tilde{M}$.

In a contact metric manifold $\tilde{M}$, the $\left( 1,1\right) $-tensor
field $h$ defined by $2h={\frak L}_{\xi }\varphi $ is symmetric and
satisfies
\begin{equation}
h\xi =0,\;\;h\varphi +\varphi h=0,\;\;\tilde{\nabla}_{X}\xi =-\varphi
X-\varphi hX,\;\;\hbox{trace}\left( h\right) =\hbox{trace}\left( \varphi
h\right) =0.  \label{cont-h}
\end{equation}
The $\left( \kappa ,\mu \right) $-{\em nullity distribution} of a contact
metric manifold $\tilde{M}$ is a distribution
\begin{eqnarray*}
N\left( \kappa ,\mu \right) \! &:&\!p\!\rightarrow \!N_{p}\left( \kappa ,\mu
\right) \!=\!\left\{ Z\in T_{p}M\mid \tilde{R}\left( X,Y\right) Z=\kappa
\left( \left\langle Y,Z\right\rangle X-\left\langle X,Z\right\rangle
Y\right) \right. \\
&&\qquad \qquad \qquad \qquad \qquad \left. +\mu \left( \left\langle
Y,Z\right\rangle hX-\left\langle X,Z\right\rangle hY\right) \right\} ,
\end{eqnarray*}
where $\kappa $ and $\mu $ are constants. If $\xi \in N\left( \kappa ,\mu
\right) $, $\tilde{M}$ is called a $\left( \kappa ,\mu \right) $-{\em %
contact metric manifold}. Since in a $\left( \kappa ,\mu \right) $-contact
metric manifold one has $h^{2}$ $=$ $\left( \kappa -1\right) \varphi ^{2}$,
therefore $\kappa \leq 1$ and if $\kappa =1$ then the structure is Sasakian.
Characteristic examples of non-Sasakian $\left( \kappa ,\mu \right) $%
-contact metric manifolds are the tangent sphere bundles of Riemannian
manifolds of constant sectional curvature not equal to one. For more details
we refer to \cite{blair02} and \cite{koufog}.

The sectional curvature $\tilde{K}\left( X,\varphi X\right) $ of a plane
section spanned by a unit vector $X$ orthogonal to $\xi $ is called a $%
\varphi $-{\em sectional curvature}. If the $\left( \kappa ,\mu \right) $%
-contact metric manifold $\tilde{M}$ has constant $\varphi $-sectional
curvature $c$ then it is called a $\left( \kappa ,\mu \right) ${\em -contact
space form} and is denoted by $\tilde{M}\left( c\right) $. The curvature
tensor of $\tilde{M}\left( c\right) $ is given by (\cite{koufog})
\begin{eqnarray}
&&\tilde{R}\left( X,Y\right) Z=\frac{c+3}{4}\left\{ \left\langle
Y,Z\right\rangle X-\left\langle X,Z\right\rangle Y\right\}  \nonumber \\
&&+\ \frac{c-1}{4}\left\{ 2\left\langle X,\varphi Y\right\rangle \varphi
Z+\left\langle X,\varphi Z\right\rangle \varphi Y-\left\langle Y,\varphi
Z\right\rangle \varphi X\right\}  \nonumber \\
&&+\ \frac{c+3-4\kappa }{4}\left\{ \eta (X)\eta (Z)Y-\eta (Y)\eta
(Z)X+\left\langle X,Z\right\rangle \eta (Y)\xi -\left\langle
Y,Z\right\rangle \eta (X)\xi \right\}  \nonumber \\
&&+\ \frac{1}{2}\left\{ \left\langle hY,Z\right\rangle hX-\left\langle
hX,Z\right\rangle hY+\left\langle \varphi hX,Z\right\rangle \varphi
hY-\left\langle \varphi hY,Z\right\rangle \varphi hX\right\}  \nonumber \\
&&+\left\langle \varphi Y,\varphi Z\right\rangle hX-\left\langle \varphi
X,\varphi Z\right\rangle hY+\left\langle hX,Z\right\rangle \varphi
^{2}Y-\left\langle hY,Z\right\rangle \varphi ^{2}X  \nonumber \\
&&+\mu \left\{ \eta (Y)\eta (Z)hX-\eta (X)\eta (Z)hY+\left\langle
hY,Z\right\rangle \eta (X)\xi -\left\langle hX,Z\right\rangle \eta (Y)\xi
\right\}  \label{sect-curv}
\end{eqnarray}
for all $X,Y,Z\in T\tilde{M}$, where $c+2\kappa =-1=$ $\kappa -\mu $ if $%
\kappa <1$.

The Riemannian invariants are the intrinsic characteristics of a Riemannian
manifold. The scalar curvature is the most studied scalar valued Riemannian
invariant on Riemannian manifolds. Let $M$ be an $n$-dimensional Riemannian
manifold. The scalar curvature $\tau $ at $p$ is given by $\tau
=\sum_{i<j}K_{ij}$, where $K_{ij}$ is the sectional curvature of the plane
section spanned by $e_{i}$ and $e_{j}$ at $p\in M$ for any orthonormal basis
$\{e_{1},\ldots ,e_{n}\}$ for $T_{p}M$. We denote by $K(\pi )$ the sectional
curvature of $M$ for a plane section $\pi $ in $T_{p}M$.

Let $M$ be an $n$-dimensional submanifold in a manifold $\tilde{M}$ equipped
with a Riemannian metric $\left\langle \,,\right\rangle $. The Gauss and
Weingarten formulae are given respectively by $\tilde{\nabla}_{X}Y=\nabla
_{X}Y+\sigma \left( X,Y\right) $ and $\tilde{\nabla}_{X}N=-A_{N}X+\nabla
_{X}^{\perp }N$ for all $X,Y\in TM$ and $N\in T^{\perp }M$, where $\tilde{%
\nabla}$, $\nabla $ and $\nabla ^{\perp }$ are Riemannian, induced
Riemannian and induced normal connections in $\tilde{M}$, $M$ and the normal
bundle $T^{\perp }M$ of $M$ respectively, and $\sigma $ is the second
fundamental form related to the shape operator $A_{N}$ in the direction of $%
N $ by $\left\langle \sigma \left( X,Y\right) ,N\right\rangle =\left\langle
A_{N}X,Y\right\rangle $. Then, the Gauss equation is given by
\begin{equation}
\tilde{R}(X,Y,Z,W)\!=\!R(X,Y,Z,W)\!-\!\left\langle \sigma \left( X,W\right)
,\sigma \left( Y,Z\right) \!\right\rangle \!+\!\left\langle \sigma \left(
X,Z\right) ,\sigma \left( Y,W\right) \!\right\rangle  \label{gauss-eqn}
\end{equation}
for all $X,Y,Z,W\in TM$, where $\tilde{R}$ and $R$ are the curvature tensors
of $\tilde{M}$ and $M$ respectively. The mean curvature vector $H$ is
expressed by $nH=\hbox{trace}\left( \sigma \right) $. The submanifold $M$ is
{\em totally geodesic} in $\tilde{M}$ if $\sigma =0$, and {\em minimal} if $%
H=0$. If $\sigma \left( X,Y\right) $ $=$ $\left\langle X,Y\right\rangle H$
for all $X,Y\in TM$, then $M$ is {\em totally umbilical}.

\section{Certain basic {\bf inequalities\label{sect-ineq}}}

We recall the following Chen's Lemma for later use.

\begin{lem}
\label{chen-lemma}{\em (\cite{chen93})} If $a_{1},\ldots ,a_{n},a_{n+1}$ are
$n+1$ $\left( n>1\right) $ real numbers such that
\[
\frac{1}{n-1}\left( \sum_{i=1}^{n}a_{i}\right)
^{2}=\sum_{i=1}^{n}a_{i}^{2}+a_{n+1},
\]
then $2a_{1}a_{2}\geq a_{n+1}$, with equality holding if and only if $%
a_{1}+a_{2}=a_{3}=\cdots =a_{n}$.
\end{lem}

For a vector field $X$ on a submanifold $M$ of an almost contact metric
manifold $\tilde{M}$, let $PX$ be the tangential part of $\varphi X$. Thus, $%
P$ is an endomorphism of the tangent bundle of $M$ and satisfies $%
\left\langle X,PY\right\rangle =-\left\langle PX,Y\right\rangle $. Let $\pi
\subset T_{p}M$ be a plane section spanned by an orthonormal basis $\left\{
e_{1},e_{2}\right\} $. Then, $\alpha (\pi )$ given by
\[
\alpha (\pi )=\left\langle e_{1},Pe_{2}\right\rangle ^{2}
\]
is a real number in the closed unit interval $\left[ 0,1\right] $, which is
independent of the choice of the orthonormal basis $\left\{
e_{1},e_{2}\right\} $. Let $\xi \in TM$ and put
\[
\beta (\pi )=(\eta (e_{1}))^{2}+(\eta (e_{2}))^{2},
\]
\[
\gamma \left( \pi \right) =\eta (e_{1})^{2}\left\langle
h^{T}e_{2},e_{2}\right\rangle +\eta (e_{2})^{2}\left\langle
h^{T}e_{1},e_{1}\right\rangle -2\eta (e_{1})\eta (e_{2})\left\langle
h^{T}e_{1},e_{2}\right\rangle .
\]
Then, $\beta (\pi )$ and $\gamma \left( \pi \right) $ are also real numbers
and do not depend on the choice of the orthonormal basis $\left\{
e_{1},e_{2}\right\} $. Of course, $\beta (\pi )\in \left[ 0,1\right] $.
\medskip

In view of (\ref{sect-curv}) and (\ref{gauss-eqn}) we state the following
Lemma.

\begin{lem}
\label{lem-tau-H-sig}In an $n$-dimensional submanifold $M$ in a $\left(
\kappa ,\mu \right) $-contact space form $\tilde{M}\left( c\right) $ such
that $\xi \in TM$, the scalar curvature and the squared mean curvature
satisfy
\begin{eqnarray}
2\tau &=&\frac{1}{4}\left\{ n\left( n-1\right) \left( c+3\right) +3\left(
c-1\right) \left\| P\right\| ^{2}-2\left( n-1\right) \left( c+3-4\kappa
\right) \right\}  \nonumber \\
&&+\frac{1}{2}\left\{ \left\| \left( \varphi h\right) ^{T}\right\|
^{2}-\left\| h^{T}\right\| ^{2}-\left( \hbox{trace}\left( \left( \varphi
h\right) ^{T}\right) \right) ^{2}+\left( \hbox{trace}\left( h^{T}\right)
\right) ^{2}\right\}  \nonumber \\
&&+2\left( \mu +n-2\right) \hbox{trace}\left( h^{T}\right) +n^{2}\left\|
H\right\| ^{2}-\left\| \sigma \right\| ^{2},  \label{tau-H-sig}
\end{eqnarray}
where
\[
\left\| \sigma \right\| ^{2}\!=\!\sum_{i,j=1}^{n}\left\langle \sigma
\!\left( e_{i},e_{j}\right) ,\sigma \!\left( e_{i},e_{j}\right)
\right\rangle ,\ \left\| Q\right\| ^{2}\,\!=\!\sum_{i,j=1}^{n}\left\langle
e_{i},Qe_{j}\right\rangle ^{2},\ Q\!\in \!\left\{ P,\left( \varphi h\right)
^{T},h^{T}\right\}
\]
and $\left( \varphi h\right) ^{T}X$ and $h^{T}X$ are the tangential parts of
$\varphi hX$ and $hX$ respectively for $X\in TM$.
\end{lem}

The equation (\ref{tau-H-sig}) is of fundamental importance and will play
main role to establish several inequalities. \medskip

Now, we prove the following contact version of Theorem~3 of \cite{chen96}.

\begin{th}
\label{th-bi-1}Let $M$ be an $n$-dimensional $\left( n\geq 3\right) $
submanifold isometrically immersed in a $\left( 2m+1\right) $-dimensional $%
\left( \kappa ,\mu \right) $-contact space form $\tilde{M}\left( c\right) $
such that $\xi \in TM$. Then, for each point $p\in M$ and each plane section
$\pi \subset T_{p}M$, we have
\begin{eqnarray}
&&\tau -K(\pi )\leq \frac{n^{2}\left( n-2\right) }{2\left( n-1\right) }%
\left\| H\right\| ^{2}+\frac{1}{8}n\left( n-3\right) \left( c+3\right)
+\left( n-1\right) \kappa  \nonumber \\
&&+\ \frac{3\left( c-1\right) }{8}\left\{ \left\| P\right\| ^{2}-2\alpha
\left( \pi \right) \right\} +\frac{1}{4}\left( c+3-4\kappa \right) \beta
\left( \pi \right) -\left( \mu -1\right) \gamma \left( \pi \right)  \nonumber
\\
&&-\ \frac{1}{2}\left\{ 2\hbox{trace}\left( h|_{\pi }\right) +\det \left(
h|_{\pi }\right) -\det \left( (\varphi h)|_{\pi }\right) \right\} +\left(
\mu +n-2\right) \hbox{trace}\left( h^{T}\right)  \nonumber \\
&&+\ \frac{1}{4}\left\{ \left\| \left( \varphi h\right) ^{T}\right\|
^{2}-\left\| h^{T}\right\| ^{2}-\left( \hbox{trace}\left( \left( \varphi
h\right) ^{T}\right) \right) ^{2}+\left( \hbox{trace}\left( h^{T}\right)
\right) ^{2}\right\} .  \label{bi}
\end{eqnarray}
The equality in $\left( \ref{bi}\right) $ holds at $p\in M$ if and only if
there exist an orthonormal basis $\left\{ e_{1},\ldots ,e_{n}\right\} $ of $%
T_{p}M$ and an orthonormal basis $\left\{ e_{n+1},\ldots ,e_{2m+1}\right\} $
of $T_{p}^{\perp }M$ such that {\bf (a)} $\pi =\hbox{Span}\left\{
e_{1},e_{2}\right\} $ and {\bf (b)} the forms of shape operators $%
A_{r}\equiv A_{e_{r}}$, $r=n+1,\ldots ,2m+1$, become
\begin{equation}
A_{n+1}=\left(
\begin{array}{ccc}
a & 0 & 0 \\
0 & b & 0 \\
0 & 0 & \left( a+b\right) I_{n-2}
\end{array}
\right) ,  \label{shape-1}
\end{equation}
\begin{equation}
A_{r}=\left(
\begin{array}{ccc}
c_{r} & d_{r} & 0 \\
d_{r} & -c_{r} & 0 \\
0 & 0 & 0_{n-2}
\end{array}
\right) ,\qquad r=n+2,\ldots ,2m+1.  \label{shape-2}
\end{equation}
\end{th}

\noindent {\bf Proof. }Let $\pi \subset T_{p}M$ be a plane section. Choose
an orthonormal basis $\{e_{1},e_{2},\ldots ,e_{n}\}$ for $T_{p}M$ and $%
\{e_{n+1},\ldots ,e_{2m+1}\}$ for the normal space $T_{p}^{\bot }M$ at $p$
such that $\pi =\hbox{Span}\left\{ e_{1},e_{2}\right\} $ and the mean
curvature vector $H$ is in the direction of the normal vector to $e_{n+1}$.
We rewrite (\ref{tau-H-sig}) as
\begin{equation}
\frac{1}{n-1}\left( \sum_{i=1}^{n}\sigma _{ii}^{n+1}\right)
^{2}\!=\!\sum_{i=1}^{n}\left( \sigma _{ii}^{n+1}\right) ^{2}\!+\!\sum_{i\not%
{=}j}\left( \sigma _{ij}^{n+1}\right) ^{2}\!+\!\sum_{r={n+2}%
}^{2m+1}\sum_{i,j=1}^{n}\left( \sigma _{ij}^{r}\right) ^{2}\!+\rho ,
\label{bi-1}
\end{equation}
where
\begin{eqnarray}
&&\rho =2\tau -\frac{n^{2}\left( n-2\right) }{n-1}\left\| H\right\|
^{2}-2\left( \mu +n-2\right) \hbox{trace}\left( h^{T}\right)  \nonumber \\
&&-\frac{1}{4}\left\{ n\left( n-1\right) \left( c+3\right) +3\left(
c-1\right) \left\| P\right\| ^{2}-2\left( n-1\right) \left( c+3-4\kappa
\right) \right\}  \nonumber \\
&&-\frac{1}{2}\left\{ \left\| \left( \varphi h\right) ^{T}\right\|
^{2}-\left\| h^{T}\right\| ^{2}-\left( \hbox{trace}\left( \left( \varphi
h\right) ^{T}\right) \right) ^{2}+\left( \hbox{trace}\left( h^{T}\right)
\right) ^{2}\right\}  \label{rho}
\end{eqnarray}
and $\sigma _{ij}^{r}=\left\langle \sigma \left( e_{i},e_{j}\right)
,e_{r}\right\rangle ,\;i,j\in \left\{ 1,\ldots ,n\right\} ;\;r\in \left\{
n+1,\ldots ,2m+1\right\} $. Now, applying Lemma~\ref{chen-lemma} to (\ref
{bi-1}), we obtain
\begin{equation}
2\sigma _{11}^{n+1}\sigma _{22}^{n+1}\geq \rho +\sum_{i\neq j}\left( \sigma
_{ij}^{n+1}\right) ^{2}+\sum_{r=n+2}^{2m+1}\sum_{i,j=1}^{n}\left( \sigma
_{ij}^{r}\right) ^{2}.  \label{bi-2}
\end{equation}
From equation (\ref{sect-curv}) and (\ref{gauss-eqn}) it also follows that
\begin{eqnarray}
K\left( \pi \right) &=&\frac{1}{4}\left\{ 3+c+3\left( c-1\right) \alpha
\left( \pi \right) -\left( c+3-4\kappa \right) \beta \left( \pi \right)
+4\left( \mu -1\right) \gamma \left( \pi \right) \right\}  \nonumber \\
&&+\frac{1}{2}\left\{ 2\hbox{trace}\left( h|_{\pi }\right) +\det \left(
h|_{\pi }\right) -\det \left( (\varphi h)|_{\pi }\right) \right\}  \nonumber
\\
&&+\sigma _{11}^{n+1}\sigma _{22}^{n+1}-\left( \sigma _{12}^{n+1}\right)
^{2}+\sum_{r=n+2}^{2m+1}\left( \sigma _{11}^{r}\sigma _{22}^{r}-\left(
\sigma _{12}^{r}\right) ^{2}\right) .  \label{bi-3}
\end{eqnarray}
Thus, from (\ref{bi-2}) and (\ref{bi-3}) we have
\begin{eqnarray}
K(\pi ) &\geq &\frac{1}{4}\left\{ 3+c+3\left( c-1\right) \alpha \left( \pi
\right) -\left( c+3-4\kappa \right) \beta \left( \pi \right) +4\left( \mu
-1\right) \gamma \left( \pi \right) \right\}  \nonumber \\
&&+\frac{1}{2}\left\{ 2\hbox{trace}\left( h|_{\pi }\right) +\det \left(
h|_{\pi }\right) -\det \left( (\varphi h)|_{\pi }\right) \right\} +\frac{1}{2%
}\rho  \nonumber \\
&&\quad +\sum_{r=n+1}^{2m+1}\sum_{j>2}\{(\sigma _{1j}^{r})^{2}+(\sigma
_{2j}^{r})^{2}\}+\frac{1}{2}\sum_{i\not{=}j>2}(\sigma _{ij}^{n+1})^{2}
\nonumber \\
&&\quad +\frac{1}{2}\sum_{r=n+2}^{2m+1}\sum_{i,j>2}(\sigma _{ij}^{r})^{2}+%
\frac{1}{2}\sum_{r=n+2}^{2m+1}(\sigma _{11}^{r}+\sigma _{22}^{r})^{2}.
\label{bi-4}
\end{eqnarray}
In view of (\ref{rho}) and (\ref{bi-4}), we get (\ref{bi}).

If the equality in (\ref{bi}) holds, then the inequalities given by (\ref
{bi-2}) and (\ref{bi-4}) become equalities. In this case, we have
\begin{equation}
\begin{array}[b]{l}
\sigma _{1j}^{n+1}=0,\ \sigma _{2j}^{n+1}=0,\ \sigma _{ij}^{n+1}=0,\quad
i\neq j>2;\medskip \\
\sigma _{1j}^{r}=\sigma _{2j}^{r}=\sigma _{ij}^{r}=0,\ r=n+2,\ldots
,2m+1;\quad i,j=3,\ldots ,n;\medskip \\
\sigma _{11}^{n+2}+\sigma _{22}^{n+2}=\cdots =\sigma _{11}^{2m+1}+\sigma
_{22}^{2m+1}=0.
\end{array}
\label{bi-6}
\end{equation}
Now, we choose $e_{1}$ and $e_{2}$ so that $\sigma _{12}^{n+1}=0$. Applying
Lemma~\ref{chen-lemma} we also have
\begin{equation}
\sigma _{11}^{n+1}+\sigma _{22}^{n+1}=\sigma _{33}^{n+1}=\cdots =\sigma
_{nn}^{n+1}.  \label{bi-7}
\end{equation}
Thus, choosing a suitable orthonormal basis $\left\{ e_{1},\ldots
,e_{2m+1}\right\} $, the shape operator of $M$ becomes of the form given by (%
\ref{shape-1}) and (\ref{shape-2}). The converse is easy to follow. $\square
\medskip $

For a submanifold $M$ of an almost contact metric manifold tangential to the
structure vector field $\xi $, we write the orthogonal direct decomposition $%
TM={\cal D}\oplus \left\{ \xi \right\} $. Moreover, if the
ambient manifold is contact also, then
\begin{equation}
\nabla _{\xi }\xi =0\quad \hbox{and\quad }\sigma \left( \xi ,\xi \right) =0.
\label{sigma-xi-xi}
\end{equation}

Next, we prove the following theorem.

\begin{th}
\label{th-bi'-1}Let $M$ be an $n$-dimensional $\left( n\geq 3\right) $
submanifold isometrically immersed in a $\left( 2m+1\right) $-dimensional $%
\left( \kappa ,\mu \right) $-contact space form $\tilde{M}\left( c\right) $
such that $\xi \in TM$. Then, for each point $p\in M$ and each plane section
$\pi \subset {\cal D}_{p}$, we have
\begin{eqnarray}
&&\tau -K(\pi )\leq \frac{n^{2}\left( n-2\right) }{2\left( n-1\right) }%
\left\| H\right\| ^{2}+\frac{1}{8}n\left( c+3\right) \left( n-3\right)
+\left( n-1\right) \kappa  \nonumber \\
&&+\frac{3\left( c-1\right) }{8}\left\{ \left\| P\right\| ^{2}-2\alpha
\left( \pi \right) \right\} +\left( \mu +n-2\right) \hbox{trace}\left(
h^{T}\right)  \nonumber \\
&&-\frac{1}{2}\left\{ 2\hbox{trace}\left( h|_{\pi }\right) +\det \left(
h|_{\pi }\right) -\det \left( (\varphi h)|_{\pi }\right) \right\}  \nonumber
\\
&&+\frac{1}{4}\left\{ \left\| \left( \varphi h\right) ^{T}\right\|
^{2}-\left\| h^{T}\right\| ^{2}-\left( \hbox{trace}\left( \left( \varphi
h\right) ^{T}\right) \right) ^{2}+\left( \hbox{trace}\left( h^{T}\right)
\right) ^{2}\right\} .  \label{bi'}
\end{eqnarray}
The equality in $\left( \ref{bi'}\right) $ holds at $p\in M$ if and only if
there exist an orthonormal basis $\left\{ e_{1},\ldots ,e_{n}\right\} $ of $%
T_{p}M$ and an orthonormal basis $\left\{ e_{n+1},\ldots ,e_{2m+1}\right\} $
of $T_{p}^{\perp }M$ such that {\bf (a)} $e_{n}=\xi $, {\bf (b) }$\pi =%
\hbox{Span}\left\{ e_{1},e_{2}\right\} $ and {\bf (c)} the forms of shape
operators $A_{r}\equiv A_{e_{r}}$, $r=n+1,\ldots ,2m+1$, become $\left( \ref
{shape-2}\right) $ and
\begin{equation}
A_{n+1}=\left(
\begin{array}{ccc}
a & 0 & 0 \\
0 & -a & 0 \\
0 & 0 & {\large 0}_{n-2}
\end{array}
\right) .  \label{shape-1'}
\end{equation}
\end{th}

\noindent {\bf Proof. }Let $\pi \subset {\cal D}_{p}$ be a plane section at $%
p\in M$. We choose an orthonormal basis $\{e_{1},e_{2},\ldots ,e_{n}=\xi \}$
for $T_{p}M$ and $\{e_{n+1},\ldots ,e_{2m+1}\}$ for the normal space $%
T_{p}^{\bot }M$ at $p$ such that $\pi =\hbox{Span}\left\{
e_{1},e_{2}\right\} $ and the mean curvature vector $H\left( p\right) $ is
parallel to $e_{n+1}$. Using $\eta \left( e_{1}\right) =0=\eta \left(
e_{2}\right) $, we get $\beta \left( \pi \right) =0=\gamma \left( \pi
\right) $. Thus, proof of (\ref{bi'}) is similar to that of (\ref{bi}). In
equality case, using (\ref{sigma-xi-xi}), (\ref{bi-7}) becomes
\begin{equation}
\sigma _{11}^{n+1}+\sigma _{22}^{n+1}=\sigma _{33}^{n+1}=\cdots =\sigma
_{nn}^{n+1}=0,  \label{bi'-7}
\end{equation}
and thus (\ref{shape-1}) is modified to (\ref{shape-1'}). $\square \medskip $

Since in case of non-Sasakian $\left( \kappa ,\mu \right) $-contact space
form, we have $\kappa <1$, and thus $c=-2\kappa -1$ and $\mu =\kappa +1$.
Putting these values in (\ref{bi}) and (\ref{bi'}), we can have direct
corollaries to Theorems~\ref{th-bi-1} and \ref{th-bi'-1}. For example,
Corollary to Theorem~\ref{th-bi-1} is as follows.

\begin{cor}
\label{cor-bi-1}Let $M$ be an $n$-dimensional $\left( n\geq 3\right) $
submanifold isometrically immersed in a non-Sasakian $\left( \kappa ,\mu
\right) $-contact space form $\tilde{M}\left( c\right) $ such that $\xi \in
TM$. Then, for each point $p\in M$ and each plane section $\pi \subset
T_{p}M $, we have
\begin{eqnarray}
&&\tau -K(\pi )\leq \frac{n^{2}\left( n-2\right) }{2\left( n-1\right) }%
\left\| H\right\| ^{2}-\frac{1}{4}n\left( n-3\right) \left( \kappa -1\right)
+\left( n-1\right) \kappa  \nonumber \\
&&-\ \frac{3}{4}\left( \kappa +1\right) \left\| P\right\| ^{2}+\frac{1}{2}%
\left\{ 3\left( \kappa +1\right) \alpha \left( \pi \right) -\left( 3\kappa
-1\right) \beta \left( \pi \right) -2\kappa \gamma \left( \pi \right)
\right\}  \nonumber \\
&&-\ \frac{1}{2}\left\{ 2\hbox{trace}\left( h|_{\pi }\right) +\det \left(
h|_{\pi }\right) -\det \left( (\varphi h)|_{\pi }\right) \right\} +\left(
\kappa +n-1\right) \hbox{trace}\left( h^{T}\right)  \nonumber \\
&&+\ \frac{1}{4}\left\{ \left\| \left( \varphi h\right) ^{T}\right\|
^{2}-\left\| h^{T}\right\| ^{2}-\left( \hbox{trace}\left( \left( \varphi
h\right) ^{T}\right) \right) ^{2}+\left( \hbox{trace}\left( h^{T}\right)
\right) ^{2}\right\} .  \label{cor-bi-1-eqn}
\end{eqnarray}
The equality in $\left( \ref{cor-bi-1-eqn}\right) $ holds at $p\in M$ if and
only if there exist an orthonormal basis $\left\{ e_{1},\ldots
,e_{n}\right\} $ of $T_{p}M$ and an orthonormal basis $\left\{
e_{n+1},\ldots ,e_{2m+1}\right\} $ of $T_{p}^{\perp }M$ such that {\bf (a)} $%
\pi =\hbox{Span}\left\{ e_{1},e_{2}\right\} $ and {\bf (b)} the shape
operators $A_{r}\equiv A_{e_{r}}$, $r=n+1,\ldots ,2m+1$ are of forms given
by $\left( \ref{shape-1}\right) $ and $\left( \ref{shape-2}\right) $.
\end{cor}

If $\kappa =1$, the $\left( \kappa ,\mu \right) $-contact space form reduces
to Sasakian space form $\tilde{M}\left( c\right) $; thus $h=0$ and (\ref
{sect-curv}) becomes the equation in Theorem~7.14 of \cite{blair02}.
Therefore, Theorem~\ref{th-bi-1} and Theorem~\ref{th-bi'-1} provide Sasakian
versions as Theorem 3.2 of \cite{KK99} and \cite{carriazo99} respectively.
Now, we recall Chen's $\delta ${\em -invariant} given by
\[
\delta _{M}(p)=\tau (p)-(\inf K)(p)=\tau (p)-\inf \{K(\pi )\ |\ \pi \hbox{
is a plane section }\subset T_{p}M\},
\]
which is certainly an intrinsic character of $M$. Improving Theorem~4.1 of
\cite{KK99}, we have the following \cite{TKK}

\begin{th}
\label{th-obst}Let $M$ be an $n$-dimensional Riemannian manifold
isometrically immersed in a Sasakian space form $\tilde{M}(c)$ of constant $%
\varphi $-sectional curvature $c<1$ with the structure vector field $\xi $
tangent to $M$. $M$ satisfies Chen's basic equality
\begin{equation}
\delta _{M}=\frac{n^{2}(n-2)}{2(n-1)}\left\| H\right\| ^{2}+\frac{1}{8}%
\{(n(n-3)c+3n^{2}-n-8\},  \label{obstr}
\end{equation}
if and only if $M$ is a $3$-dimensional minimal invariant submanifold.
Hence, Chen's invariant becomes $\delta _{M}=2$.
\end{th}

\section{Invariant submanifolds\label{sect-inv-sub}}

A submanifold $M$ of an almost contact metric manifold $\tilde{M}$ with the
structure $\left( \varphi ,\xi ,\eta ,\left\langle \,,\right\rangle \right) $
is called an invariant submanifold if $\varphi T_{p}M\subset T_{p}M$. If $%
\tilde{M}$ is contact also, then $\xi \in TM$, $\sigma \left( X,\xi \right)
=0$ and $M$ is minimal (\cite{blair02}). On the other hand, we have the
following

\begin{prop}
\label{prop-tot-umb}Every totally umbilical submanifold $M$ of a contact
metric manifold such that $\xi \in TM$ is minimal and consequently totally
geodesic.
\end{prop}

The proof follows from $H=\left\langle \xi ,\xi \right\rangle H=\sigma (\xi
,\xi )=0$, where (\ref{sigma-xi-xi}) is used. Choosing an orthonormal basis $%
\left\{ e_{i},\varphi e_{i},\xi \right\} $, $i=1,\ldots ,\frac{n-1}{2}$, we
also can prove

\begin{prop}
\label{prop-tr-h-phi-h}In an $n$-dimensional invariant submanifold of a
contact metric manifold, we have
\begin{equation}
\left\| P\right\| ^{2}\!=\!n-1\hbox{,\ }\hbox{trace}\left( h^{T}\right) \!=\!%
\hbox{trace}\left( \left( \varphi h\right) ^{T}\right) \!=\!0,\;\left\|
\left( \varphi h\right) ^{T}\right\| ^{2}\!=\!\left\| h^{T}\right\| ^{2}.
\label{inv-tr-h=0}
\end{equation}
\end{prop}

Thus, for an $n$-dimensional invariant submanifold in a $\left( \kappa ,\mu
\right) $-contact space form $\tilde{M}\!\left( c\right) $, the scalar
curvature and the second fundamental form satisfy
\begin{equation}
2\tau =\frac{1}{4}\left( n-1\right) \left\{ \left( n+1\right) c+8\kappa
+3n-9\right\} -\left\| \sigma \right\| ^{2}.  \label{inv-tau-sigma}
\end{equation}

In view of the above equation, we can state the following theorem.

\begin{th}
\label{th-inv-tau-sig}For an $n$-dimensional invariant submanifold
isometrically immersed in a $\left( \kappa ,\mu \right) $-contact space form
$\tilde{M}\left( c\right) $, we get
\begin{equation}
\tau \leq \frac{1}{8}\left( n-1\right) \left\{ \left( n+1\right) c+8\kappa
+3n-9\right\}  \label{inv-tau-sig}
\end{equation}
with equality if and only if the invariant submanifold is totally umbilical,
where $c=-2\kappa -1$ if $\kappa <1$.
\end{th}

For each point $p\in M$, we put (\cite{carriazo99})
\begin{equation}
\delta _{M}^{{\cal D}}\left( p\right) =\tau \left( p\right) -\inf \left\{
K\left( \pi \right) :\hbox{plane sections\ }\pi \subset {\cal D}_{p}\right\}
.  \label{inv-delta'}
\end{equation}
Now, we conclude the paper with the following theorem.

\begin{th}
\label{th-chen-ineq}For an $n$-dimensional invariant submanifold
isometrically immersed in a $\left( \kappa ,\mu \right) $-contact space form
$\tilde{M}\left( c\right) $, we get
\begin{equation}
\delta _{M}^{{\cal D}}\leq \frac{n-3}{8}\left\{ \left( n+3\right) c+3\left(
n-1\right) \right\} +\left( n-1\right) \kappa .  \label{inv-chen-ineq}
\end{equation}
\end{th}

\noindent {\bf Proof.} Let $\pi \subset {\cal D}_{p}$ be a plane section at $%
p\in M$. We can choose unit vectors $e$ and $Pe$ such that $\pi =\hbox{Span}%
\left\{ e,Pe\right\} $. Thus, we get $\alpha \left( \pi \right) =1$, $%
\hbox{trace}\left( h|_{\pi }\right) =0$ and $\det \left( h|_{\pi }\right)
=\det \left( (\varphi h)|_{\pi }\right) $. Using these facts along with (\ref
{inv-tr-h=0}) in (\ref{bi'}), we get (\ref{inv-chen-ineq}). $\square $

\noindent Mukut Mani Tripathi\newline
Department of Mathematics and Astronomy, \newline
Lucknow University, \newline
Lucknow 226 007, India.\newline
Current Address\newline
Department of Mathematics, \newline
Chonnam National University, \newline
Kwangju 500-757, Korea.\newline
email:\quad {\tt mm\_tripathi@hotmail.com}\bigskip \newline
Jeong-Sik Kim\newline
Department of Mathematics Education, \newline
Sunchon National University, \newline
Chonnam 540-742, Korea.\newline
email:\quad {\tt jskim01@hanmir.com}\bigskip \newline
Jaedong Choi\newline
Department of Mathematics, \newline
P.O. Box 335-2, Air Force Academy, \newline
Ssangsu, Namil, Cheongwon, \newline
Chungbuk, 363-849, Korea. \newline
e-mail:\quad {\tt jdong@afa.ac.kr}

\end{document}